# Long term stability and reproducibility of magnetic colloids are key issues for steady values of Specific Power Absorption through time.

B. Sanz* [a,b,c], M.P. Calatayud [a,b], N. Cassinelli [c], M.R. Ibarra [a,b] and G. F. Goya* [a,b]

**Abstract:** Virtually all clinical applications of magnetic nanoparticles (MNPs) require the formulation of biocompatible, water-based magnetic colloids. For magnetic hyperthermia, the requirements also include a high colloidal stability against precipitation and agglomeration of the constituent MNPs, in order to keep the heating efficiency of the ferrofluid in the long term. The specific power absorption (SPA) of single-domain MNPs depends critically on the average particle size and size distribution width, therefore first-rate reproducibility among different batches regarding these parameters are also needed. We have studied the evolution of the SPA of highly reproducible and stable water-based colloid composed of polymer-coated $Fe_3O_4$ magnetic nanoparticles. By measuring the specific power absorption (SPA) values along one year as a function of field amplitude and frequencies (H ≤ 24 kA/m; 260 ≤ *f* ≤ 830 kHz), we demonstrated that SPA in these samples can be made reproducible between successive synthesis, and stable along several months, due to the *in situ* polymer coating that provides colloidal stability and keeps dipolar interactions negligible.

## Introduction

Due to the irruption of the nanoparticle-containing products for human uses, there is a growing body of statutory requirements related to the safety of nanomaterials for public health. These requirements are the basis of a regulatory framework seeking to develop "robust methods in identifying and characterizing nanomaterials in consumer products; to understand their effects on human exposure; and to develop scientists to advance nanomaterials in consumer product safety research".[1] For clinical uses the specifications are even more exigent, and for nanoparticles (usually in colloidal form) a major requirement is the quality control of the manufacturing process. One of such evolving materials are the magnetic nanoparticles (MNPs) used in clinical protocols of magnetic resonance imaging or magnetic fluid hyperthermia. For the latter, the MNPs are used as heating agents for thermal ablation of tumours[2], by applying an alternating magnetic field (AMF) and transforming magnetic energy into heat.[3]

For biomedical applications, the MNPs must be suspended in a biocompatible liquid (i.e. water) constituting a magnetic colloid.[4] The MNPs dispersed in a liquid carrier are usually made of iron oxide (often $\gamma$-$Fe_2O_3$ or $Fe_3O_4$) and functionalized with organic coatings that depend on the application needed. Synthetic routes such as co-precipitation,[5],[6] high-temperature decomposition[3] or mild oxidative hydrolysis[5, 7] are readily available to [8] produce these kinds of materials in semi-industrial quantities. Some of these routes produce MNPs with very small degree of polydispersity (with relative values of 10% or less) such as the high-temperature decomposition.[9],[10] Since power absorption depends critically on particle volume, these chemical routes provide good samples for testing the current SPA models.

Bottom line is, for clinical applications of MNPs as heating agents, it is not the control of size homogeneity what matters, but the control of reproducibility from batch to batch. The reason is that therapeutic applications require that a given material be able to deliver the same amount of heat under fixed field conditions (amplitude and frequency). Therefore it is the reproducibility of the <d> and σ parameters along successive synthesis what matters.

Among the synthesis routes mentioned above, the oxidative hydrolysis[7] provides many advantages for the production of heating agents. The crystalline iron oxide magnetic cores are directly obtained in aqueous media, with obvious advantages compared to organic solvents for biocompatibility. By in situ adding a polymer such as PEI or PAA, the resulting magnetic cores have a final diameter between 30 and 100 nm, depending on the type and size of the polymer. Indeed, it has been previously reported that the final particle size is controlled by the type of polymer used.[7]

[a] B. Sanz, M.P. Calatayud, M.R. Ibarra and G.F. Goya
Instituto de Nanociencia de Aragón (INA)
Universidad de Zaragoza
Edificio I+D, Calle Mariano Esquillo s/n (Campus Río Ebro), 50018, Zaragoza, Spain.
E-mail: beasanz@unizar.es, goya@unizar.es.
[b] B. Sanz, M.P. Calatayud, M.R. Ibarra and G.F. Goya
Departamento de Física de la Materia Condensada
Universidad de Zaragoza
Calle Pedro Cerbuna 12, 50009, Zaragoza, Spain.
[c] B. Sanz and N. Cassinelli.
nB Nanoscale Biomagnetics S.L.
Calle Panamá 2, Zaragoza 50012, Spain.

The specific power absorption values (SPA) of a heating colloid usually changes over time due to their tendency to aggregate.[11],[12] Since magnetic cores are single-domain particles, thermodynamic considerations show that the magnetic moments tend to aggregate to form structures to lower their energy state.[13] Therefore, colloidal stability against agglomeration and precipitation is also an essential requirement for biomedical applications of ferrofluids.[14],[15] It is mostly the surface chemistry that determines the trend of the nanoparticles in the media carrier to aggregate in clusters, due to magnetic attractive forces between the particles and to the Van-der Walls long range attractive forces between particles.[13] The balance between several forms of energy (attractive and repulsive forces) and the thermal energy which determine the stability is influenced by the volume fraction, size distribution and temperature.[4] In magnetic colloids the extra presence of attractive-dipolar magnetic interactions between particles can, depending on the average particle size, accelerate the agglomeration process and hence lower stability and cause precipitation. Equally long range repulsive forces are required to counteract these attractive interactions. Coating the particles with long chain polymer, a steric hindrance between the particles is given to avoid the aggregation.[13] The expression for the power absorbed for an ensemble of MNPs of initial dc susceptibility $\chi_0$, under an AMF of amplitude $H_0$ and frequency $f$ is given by [16]

$$P = \chi_0 H_0^2 \mu_0 \pi \frac{2\pi f^2 \tau}{1+(2\pi f \tau)^2} \quad \text{Eq. 1}$$

where $\mu_0$ is the permeability of the free space and $\tau$ is the relaxation time of the magnetic moments. The main dissipation mechanisms by which the magnetic moments within each single domain MNPs can rotate under the action of an external magnetic field are: a) the Brownian relaxation, which is the physical rotation of the MNPs against viscous torques from the surrounding media (e.g., carrier fluid or tissue), and b) the Néel relaxation due to the rotation of the atomic magnetic moments within the crystal lattice of the MNPs against an effective magnetic anisotropy, from magnetocrystalline or shape origin.[17] These relaxation mechanisms are given by the corresponding relaxation times for Brownian and Néel relaxation ($\tau_B$ and $\tau_N$, respectively), through the expressions

$$\tau_B = \frac{3\eta V_H}{k_B T} \quad \text{Eq. 2}$$

$$\tau_N = \frac{\sqrt{\pi}}{2}\tau_0 \frac{\exp\left(\frac{KV}{k_B T}\right)}{\sqrt{\frac{KV}{k_B T}}} \quad \text{Eq. 3}$$

These equations show the intrinsic dependence of relaxation times with the viscosity of the medium (η), the hydrodynamic volume ($V_H$), the Boltzmann constant ($k_B$), the temperature (T), the magnetic anisotropy (K) and core volume (V). In particular, there is an exponential dependence of the SPA with the particle volume in the case of Néel mechanisms, and also a linear dependence of Brownian mechanism with hydrodynamic size, thus colloidal parameters can influence the final SPA of the colloid. Both mechanisms take place simultaneously in the system, thus small differences on the volume particles can result in changes of the power dissipation through changes on both relaxation mechanisms.

In this work we report a systematic study on a biocompatible stable colloid of magnetic nanoparticles (MNPs) covered with polyethylene-imine (PEI), tracking the SPA of several samples along one year. The colloid was synthetized by a simple chemical route and is potentially scalable to industrial grade. We have measured the colloidal stability and the resulting SPA values at different times, correlating the observed changes to the evolution of the colloidal agglomeration and/or precipitation.

## Results and Discussion

TEM images of the PEI-MNPs were performance to get the sample's average particle size and shape. Figure 1 shows that the particles present an octahedral shape with an average particle size of 25±5 nm and the hydrodynamic diameter of $<D_H>$ = 155±25 nm which has been previously reported.[18] HRTEM image (figure 1, B) shows the crystalline lattice; furthermore the polymer layer is identified as an amorphous layer over the crystalline edge whose thickness is 2 nm. Particles size have been extracted by several particles synthesis in order to check the reproducibility of the synthesis process, results are compared in figure 1, C, where a Gaussian fit have been performed in each histograms to calculates the size average and the standard deviation (σ).

All samples synthesized by the method reported here resulted in MNPs with average particle size <d> around 25 nm, i.e., within the size window of maximum heating efficiency and thus potentially relevant for magnetic hyperthermia.[19] Also the size distribution σ obtained through successive synthesis were found within the 2 < σ < 5 nm range (i.e., 0.08 < σ/<d> < 0.2), which represents good values as compared to other synthesis routes. Indeed, MNPs synthesis through organic routes such as thermal decomposition of organic precursors, have reported size distributions among the narrowest values (σ<10%), [9] [10] [20] [21] whereas co-precipitation strategies based on the hydrolysis of a stoichiometric mixture of $Fe^{2+}$ and $Fe^{3+}$ ions to form $Fe_3O_4$ NPs usually result in broader distributions (σ>10%). [22] [23] [24] Also, reverse micelle synthesis has been reported to give average particle sizes within 2.92 < <d> < 8.95 nm (i.e., 0.061< σ/<d> <0.12)[25] Besides Gen et al. prepared α-FeOOH nanorods at room temperature by a micoremulsion approach whose size were 8.2 ± 1.5 nm (σ/<d>=0.18) in diameter and 106 ± 16 nm (σ/<d>=6.62)in length.[26]

On the other hand, the reproducibility of both the average particle size and size distributions, as observed in figure 1, are excellent for the PEI-MNPs samples. The overlapping size distributions will result in an unprecedented reproducibility of the heating performance, as discussed below. It has been already reported that the size of the MNPs can be controlled by the in-situ addition of polymer during the synthesis via the oxidative hydrolysis.[18] The mechanism is related to the electrostatic self-assembly of the PEI chains onto the $Fe_3O_4$ surface during the synthesis at high pH,[27] providing an excellent reproducibility of both the average particle size and the size dispersion. Is this property, together with the high stability of the colloidal MNPs obtained, that makes this synthesis route the best choice for biomedical applications of magnetic hyperthermia. An additional benefit of this technique is that it can be easily scalable to industrial standards. The polymer acts as stabilization agent since it easily absorbs at the magnetic nanoparticle surface, providing positively charged amino groups that stabilizes the particles by electrostatic repulsion.[28]

The narrow distribution make them potentially candidates for magnetic hyperthermia, since the size and distribution has an important impact on the heating efficacy, then a small change in particles diameter provoke huge differences in SPA. MNPs with broad ranges of sizes have different saturation magnetization values and anisotropies reducing the maximum SPA.[29]

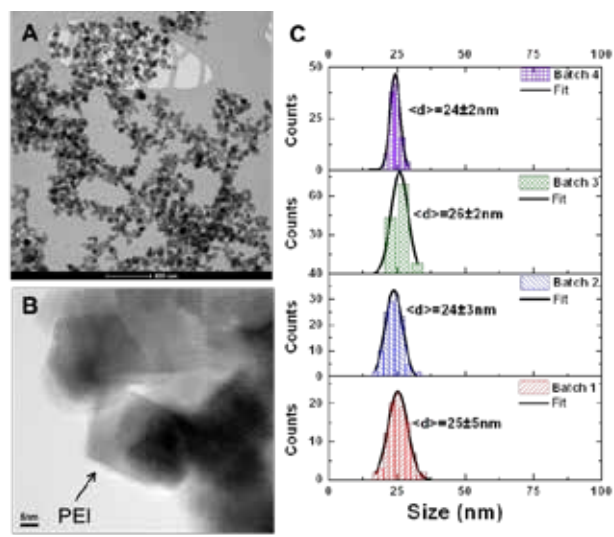

**Figure 1** A) TEM micrograph of PEI-$Fe_3O_4$ MNPs; B) selected HRTEM image of a single MNP showing the presence of the PEI-polymer layer at the surface; C) histograms and Gaussian fit curves from four different batches of MNPs-PEI, showing the reproducibility of average size and distributions.

Regarding magnetic hyperthermia applications, a strict reproducibility of the final particle size and size distribution is a key consideration, since the heat generated by an ensemble of MNPs in a magnetic colloid can substantially change by tiny shifts in the average particle size [30] [31] or size distribution.[16] These variations hinder the possibility of a standardized response of a given material, resulting in large variations of the thermal doses under fixed experimental conditions. Therefore what matters is the best reproducibility from batch to batch and not only the control of the size dispersion in a single synthesis. In this regard, oxidative hydrolysis has been already reported to produce samples with a very low variability in the final average particle sizes.[7]

A. *Magnetic Hyperthermia and SPA calculation*

The heating capability of a magnetic colloid is given by the SPA under given magnetic field amplitude and frequency. The experimental SPA value for a magnetic colloid is obtained from the calorimetric relationship[32]

$$SPA = \frac{m_{NP}c_{NP}+m_l c_l}{m_{NP}}\left(\frac{\Delta T}{\Delta t}\right)_{max} \qquad \text{Eq. 4}$$

where $c_l$ and $c_{NP}$ are the heat capacities of the solvent (water, $c_l$ = 4.18J/K $cm^3$) and MNPs, respectively; and $m_l$, $m_{NP}$ are the mass of the liquid and MNPs, respectively. The term between parenthesis ΔT/Δt is the heating rate of the sample, evaluated at the maximum slope (see below).

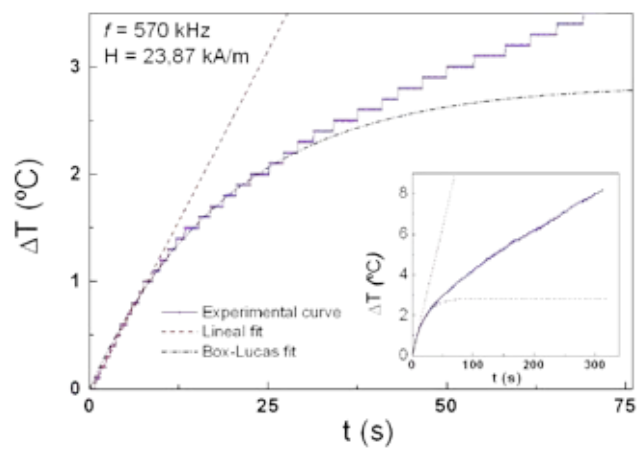

**Figure 2** Typical temperature vs. time curve for a colloidal suspension under alternate magnetic field (H=23.87 kA/m; f=570 kHz) with NPs concentration φ = 2 mg Fe$_3$O$_4$/ml. Experimental data (open circles) was fitted with a lineal dependence (dashed line) and a modified Box-Lucas function (dot-dashed curve) to obtain (dT/dt)$_{max}$.

Conceptually, a SPA experiment involves the continuous delivery of heat from the sample that is thermally insulated in an (ideally) adiabatic environment. This thermal insulation would imply a continuous, linear increase of temperature with time. In actual experimental conditions, however, adiabaticity is seldom possible, and thus heating rates are a non-linear function of time, as shown in figure 2. Therefore an appropriate model for time dependence of temperature should include heat exchange with the medium, absorption from sample holder materials yielding an asymptotic thermal equilibrium for long times. We have chosen an empirical non-linear fit procedure, making a qualitative analysis of the previous assumptions, and related to the methodology first presented by Box and Lucas in 1959 [33] for optimal design of experiments based on non-linear models. The Box-Lucas equation for the non-adiabatic temperature increase of an SPA experiment is usually expressed as [34] [35]

$$T(t) = A(1 - e^{-Bt}) + C \quad \text{Eq. 5}$$

Since parameters A and B are related with the final temperature (i.e., T(t=∞)) and the heating rate of a given sample, respectively, a more explicit form of this expression to display the relevant parameters of the heating curve is

$$T(t) = (T_0 - T_{eq})e^{-t/\tau} + T_{eq} \quad \text{Eq. 6}$$

where parameter of equations 5 and 6 are related by $C = T(t=0) = T_0$, $A + C = T(t=\infty) = T_{eq}$, and $B = 1/\tau$. In this form, the initial temperature of the sample $T_o$ (ºC) and the final equilibrium temperature $T_{eq}$ (ºC) can be directly extracted from inspection of the fitting parameters. The exponential factor τ(s) is a characteristic heating time that depends on sample properties. This modified-Box-Lucas (MBL) approximation to describe the time-dependence of temperature during power absorption in magnetic colloids has been previously reported.[35-36] The product A*B represents the initial heat rate and can be used to obtain the SPA from the derivative of Eq. 6

$$\left(\frac{dT}{dt}\right)_{max} = A \cdot B = (T_{eq} - T_o)\frac{1}{\tau} \quad \text{Eq. 7}$$

As compared to a simple linear fit of T(t) curves, the MBL fitting procedure has the practical usefulness of giving the relevant physical parameters just by inspection. In particular, the knowledge of $T_{eq} = T(t=\infty)$ allows to estimate the expected final temperature under given experimental conditions without measuring for long times, whereas the expression gives the maximum heating rate of the colloid. Therefore, combining Eq. 4 with the maximum derivative criteria (Eq. 7) the SPA can be expressed as

$$SPA = \frac{m_{NP}c_{NP} + m_l c_l}{m_{NP}}\left(\frac{dT}{dt}\right)_{max} = \frac{m_{NP}c_{NP} + m_l c_l}{m_{NP}}(A \cdot B) \quad \text{Eq. 8}$$

This equation can be further simplified assuming $m_{NP} c_{NP} \ll m_l c_l$ and defining de sample concentration as φ = $m_{NP}/V_l$ where $V_l$ is the volume (in ml) of colloid having a mass of nanoparticles $m_{NP}$ (in mg)

$$SPA = \frac{\delta_l c_l}{\varphi}\left(\frac{dT}{dt}\right)_{max} = \frac{\delta_l c_l}{\varphi}(A \cdot B) \quad \text{Eq. 9}$$

All samples studied here were conditioned to have φ ≈ 2 mg Fe$_3$O$_4$/ml.

We have compared the linear and modified Box-Lucas fits on the data extracted from a large number (>50) measurements on the samples along several months. Figure 2 shows a typical temperature curve vs. time and the resulting fits for different time windows.

The experimental curve shows a non-linear increase during the first few minutes after the AMF is turned on. After thermal profile of sample holder reaches a steady state, the temperature increases linearly. For fitting this behaviour, a linear function is usually used[30] but also a Box-Lucas function has been proposed to extract the SPA values.[35, 37] The initial increase of temperature can be extracted from $(dT/dt)_{max}$ so that a standardized protocol for calculating the SPA can be given. Depending on the function used, the $(dT/dt)_{max}$ estimations give slightly differences (see figure 2). We have compared both approaches by first using the linear fit to obtain the $(dT/dt)_{max}$ yielding $(dT/dt)_{max}$=0.127 and SPA=266 W/g $Fe_3O_4$. On the other hand, the Box-Lucas fit provided fitting parameters A=2.838 and B=0.052, resulting in a product (A.B)=0.147 and thus SPA=307 W/g $Fe_3O_4$. Therefore SPA values obtained for both adjustments differ in a small (≤5%) SPA overestimation for the modified Box-Lucas fit. The exponential shape at initial times is usually observed when measurements are done under non-adiabatic conditions. In our case the Box-Lucas fit yielded an 'extrapolated' temperature increase of $\Delta T_{max}$~2.8ºC that is much smaller than the actual temperature after several minutes. Furthermore, the linear behaviour observed at longer times also reflects the quasi-adiabatic conditions for a heated sample dissipating power at a fixed rate. This linear trend demonstrates that the thermal insulation of the sample space is good enough to consider it as quasi-adiabatic even for long times (minutes).[38]

Regarding the effect on the heating efficiency, the basic mechanisms by which the transformation from magnetic energy to heat takes place has been recently modelled under some simplifying assumptions[39] with good agreement to several experimental systems[40] [41] [3] Irrespective of the theoretical approximation to the SPA heating mechanisms, all models contains a sharp dependence with magnetic volume and dispersity of the ensemble, mainly due to the exponential dependence of the thermally assisted magnetic relaxation (Néel relaxation) with particle size.

It has been shown in the previous section that, for this synthesis route, the resulting MNPs have an average particle size around 24-26 nm, determined by the size of the PEI polymer. They are also produced with a low degree of polydispersity ($\sigma$ = 2-5 nm) that is expected to contribute to the heating efficiency of the magnetic colloid. We would like to remark that, unlike usually stated for other methods, it is not the low size dispersion what constitutes the main requirement for magnetic hyperthermia but the reproducibility of the MNPs size distribution, which grants reproducibility of SPA values and thus of 'thermal dose' under fixed experimental conditions.

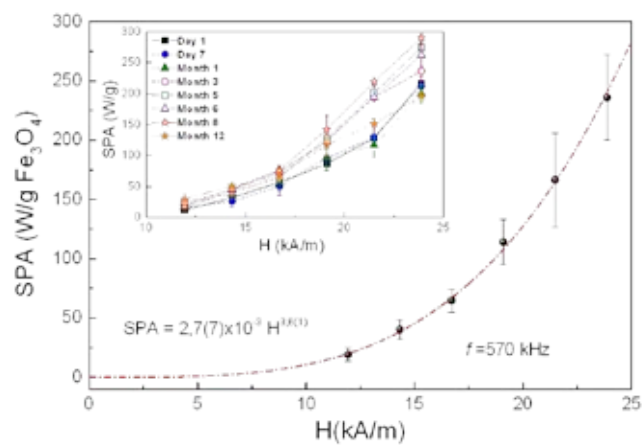

**Figure 3** Field amplitude dependence (H=0-24 kA/m, f=570 kHz) of SPA average fitted by a power equation. The inset shows the measurements for the same PEI-$Fe_3O_4$ NPs month by month.

All measurements were performed as a function of magnetic field amplitude and frequency, as described in the experimental section, and these measurements were repeated along several months to follow the evolution of the SPA with time. For each experimental session along time, the concentration of $Fe_3O_4$ was determined *just before* each measurement. All samples were initially conditioned to have a concentration of ~2 mg/ml. The time required for each measurement was about 3 minutes.

Figure 3 shows the dependence of SPA value with magnetic field H (kA/m), all measurements have been carried out systematically at *f*=570 kHz and repeated month by month at same conditions in order to track the effect of colloidal stability on SPA values. The SPA average of all the measurements could be fitted with a power law $SPA \sim H^\lambda$. The expected dependence of the SPA(H,f) curves, from the linear response theory (LRT), is a quadratic one, i.e., $SPA \sim H^\lambda$ with $\lambda$=2 (see Eq. 1).[30] It is known that the LRT is strictly valid for H << $H_C$, where $H_C$ is the coercive field of the MNPs. Given the maximum $H_0$ values applied in our experiments and the relatively small magnetic anisotropy of $Fe_3O_4$, the LRT model cannot be used. Therefore the SPA of our PEI-$Fe_3O_4$ particles is expected to follow a $\lambda \neq 2$ dependence in Eq.1. Figure 3 shows the resulting fit of experimental, which yielded a SPA ~ $H^{3.6}$. This large value of $\lambda$ = 3.6(1) has been verified in all measurements of the different batches along several months. This is the expected behaviour since the base material (i.e., the PEI-Fe$_3$O$_4$ MNPs) remained essentially unchanged along time.

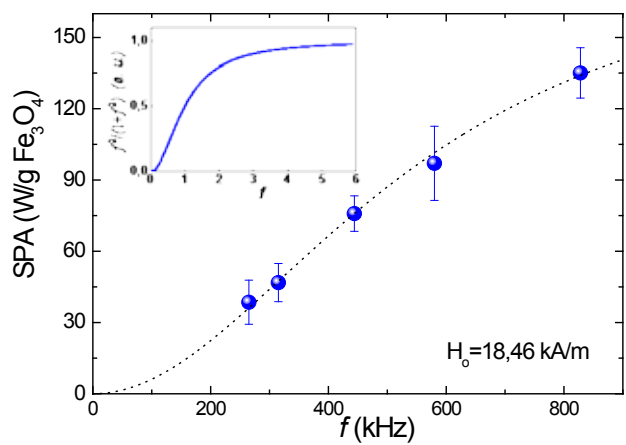

**Figure 4** Frequency dependence of *SPA* vs. frequency ($H_0$=18.46 kA/m) fitted with the expected S-shaped function (see Inset) SPA ~ $f^2/(1+f^2)$ from the simple model in ref [17]

Regarding the SPA dependence with frequency (figure 4) a linear fit is often used when analysing experimental data. However, direct inspection of Eq. 1 indicates that even if the LRT approximation is valid, this dependence is not linear, as plotted in the inset of Figure 4, yet the linear dependence should only be considered for small ranges of frequency.

A parameter named intrinsic loss power (ILP=SPA/($f \cdot H^2$) has been proposed by Kallumadil and co-workers[35] as a tentative to define an instrument-independent parameter designed to compare measurements under different amplitude and frequency conditions. As discussed above, the SPA(H,f) has not the assumed quadratic (H) and linear (f) dependences and thus the ILP approximation is expected to fail under the usual laboratory conditions. Although we cannot correlate the $\lambda$=3.6(1) value fitted from our data to any current theoretical model, to the best of our knowledge, this value should be considered as a guide for future theoretical modelling in Fe$_3$O$_4$ based single-domain MNPs.

### B. Colloidal stability through time and stillness of SPA values

The current state regarding standardized protocols in the field of magnetic hyperthermia is far from optimal. The lack of standard materials, instrumentation and protocols has its origins in many different complexities of the technique, and only recently the awareness of this deficiency has been raised.[30] From the material perspective, the capacity to produce a given value of SPA in a reproducible way is governed by the reproducibility on the average particle size and dispersion of the constituent MNPs, and in a lesser amount by the effective magnetic anisotropy. Although there are synthetic routes capable to produce nearly single-size MNPs,[9] these protocols have not shown a similar precision on reproducibility of a given <d> value from batch to batch.

Moreover, the persistence of a SPA value through time depends on the colloidal stability of the MNPs that produce the heating. Although notable improvements in colloidal stability have been reported [42] [43] the best results are reported for a particle size window of 3-10 nm, is well below the optimal size for Fe$_3$O$_4$ particles as heating agents. [30] [44] [45] Maximum values of SPA has been reported for Fe$_3$O$_4$ particles to be around 20-30 nm,[9, 46] and therefore colloidal stability is difficult to obtain because not only aggregation but also precipitation processes. For the PEI-coated MNPs of this work, colloidal stability is provided by the polymer, which also defines the electrostatic and steric repulsion between MNPs. [27]

To assess the stability of the magnetic colloids through time, we performed SPA measurements at selected and fixed conditions (*f* = 580 kHz, H= 23.87 kA/m) along twelve months. The resulting values, shown in figure 5, fluctuated between 200(10) and 290(20) W/g. In Figure 5 the evolution of two batches of colloids are displayed, and it is clear that the change of SPA through time is quite similar for the two samples, and more pronounced between different months. Whereas changes in SPA amounted a maximum value of 30% difference (between months 1 and 8) the corresponding differences between sample 1 and 2 never exceeded a 10%.

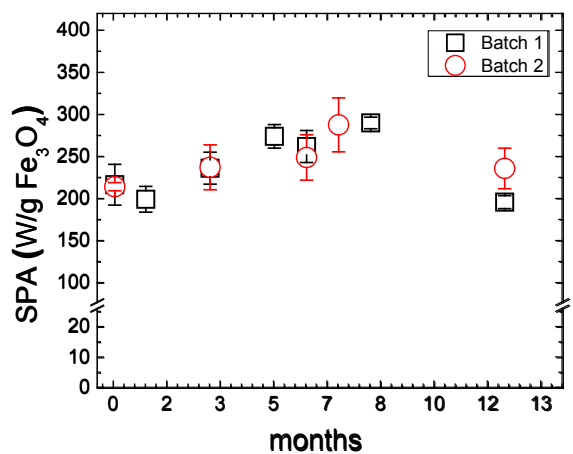

**Figure 5** *Tracking the SPA value of two different samples with time at 570 kHz and 23.87 kA/m.*

Whether these fluctuations corresponded to changes in the colloidal state of the sample required information about the change the evolution of the hydrodynamic diameter with time. For this purpose, dynamic light scattering (DLS) measurements were performed on the same day of each SPA experiment. The results plotted in Figure 6 show that the hydrodynamic sizes recorded at different months were quite variable, and we attributed this to the variable effects of the sonication previous to each measurement, resulting in different particle distributions. The evolution of the $D_H$ values, tracked for the first 30-days period, yielded no measurable changes, but increasing size of aggregates were observed at longer times, up to an average of 670 nm (month 8). However, these changes did not resulted in similar changes on SPA values. It is known that MNPs interactions can modify the effective anisotropy of the particles, and thus the dynamics of the magnetic relaxation, with the consequent changes in power absorption values. If 'nude' MNPs came into direct contact, it has been proposed that magnetic coupling through inter-particle super-exchange O-Fe-O paths can provoke major changes in the magnetic response of MNPs. However, since MNPs are often coated with different types of polymer, direct surface contact seems unlikely even within MNP-aggregates, and magnetic dipolar interactions are the dominant effects. Clearly, as the dipolar interactions depend as $1/R^3$ being R the interparticle distance, this type of interactions will decrease sharply with increasing R (i.e., the effective MNP coating thickness). For the present samples, it is clear that no major effects can be attributed to variations in aggregation states and we attributed this to the PEI content within the aggregates that keeps the dipolar interaction among magnetic cores relatively weak. Previous reports on the time evolution of PEG-coated MNPs showed that effective diameter increases due aggregation along the first days (even hours) after synthesis.[47] Kallumadil et al. published the relationship between ILP and hydrodynamic diameter indicating an optimum value to heat generation around 70 nm.[35] According with this affirmation the SPA values should be decreased with the increment of $D_H$, but they are invariable with time as could be observed in our case.

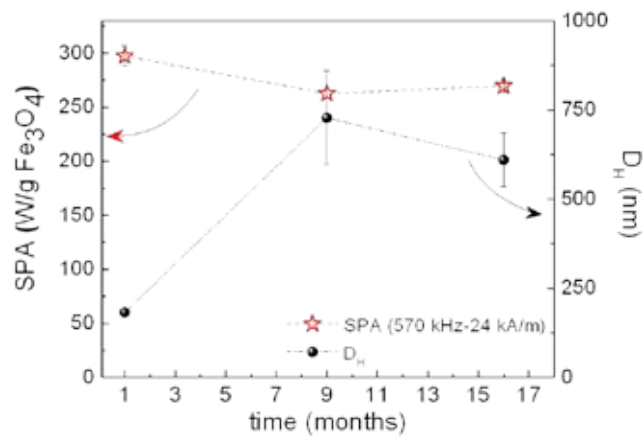

**Figure 6** Evolution of SPA (open stars) values along several months, and effective diameter $D_H$ (filled circles) as obtained from DLS. Each experimental point corresponds to the average of three different samples.

The *as prepared* samples showed $D_H$ values around 200 nm, which remained essentially unchanged during the first 2-3 months after preparation. This aggregation decreases the SPA moderately (10%, see Figure 6). Assuming that the PEI-MNPs are rigid spheres,[18] the number of particles composing these aggregates in water can be estimated to be 124 MNPs/aggregate for the *as prepared* colloids, increasing to 7950 MNPs/aggregate after several hours. Figure 7 shows a schematic representation of these aggregates formed by PEI-$Fe_3O_4$ NPs in water, where the magnetic moment of the individual particles is depicted as randomly orientated. Also, some recent calculations of both non-interacting and interacting cases in the longitudinal and random configuration for chain-chain interactions have revealed some direct effects over the magnetic hyperthermia, i.e., that the SPA of the colloids tends to decrease with increasing dipolar interactions.[48] According to this picture, it is expected that for non-interacting MNPs the Néel relaxation would remain essentially unchanged after agglomeration, provided that polymer coating prevents MNPs proximity. Thus, within large agglomerates each particle will contribute to the SPA in the same amount that a 'free' colloidal particle.

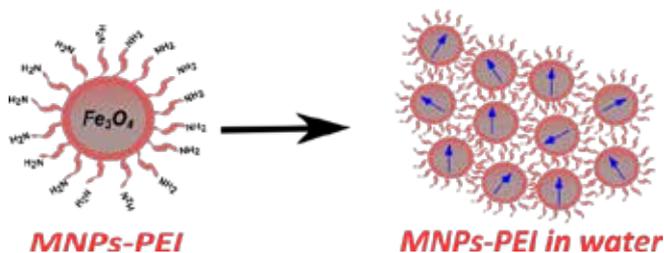

**Figure 7** Schematic representation of PEI-Fe$_3$O$_4$ NPs forming non-interacting magnetic spins aggregation in the ferrofluid.

## Conclusions

The magnetic colloid studied here provided a noticeable improvement over current standards of reproducibility in the field of magnetic hyperthermia. Given the characteristics of the PEI-coating polymer that controls the final size and size distribution, the resulting MNPs showed suitable for uses as a SPA standard for a given magnetic hyperthermia setup. Specifically, the polymer coating acts as stabilization agent through electrostatic repulsion of positive charges among NH$_4^+$ groups. Collateral advantages include cheapness and scalability of the synthesis process.

The stability of these colloids and their reproducibility batch to batch has been the key factor to study the power dissipation of this colloid for more than one year.

The average particle size and narrow size distribution (<d>=25(5) nm) yielded SPA values up to 239(32) W/g (*f*=570 kHz, H=23.87 kA/m), in agreement with prevailing models on magnetic relaxation on MNPs. Although the SPA values could be improved by narrowing the size dispersion, this water-based, one-step synthesis route granted an unprecedented level of reproducibility from batch to batch, to the best of our knowledge, as well as reproducible SPA values through time. The evolution of the SPA values has been studied as a function of magnetic field amplitude and frequency. The experimental value of the exponent $\lambda$ =3.6(1) in the amplitude-dependence SPA~H$^\lambda$ reported here should be considered as a guide for future modelling of these mechanisms in magnetite MNPs. The experimental stability of SPA values with time observed along 12 months is explained by the contribution of a) the colloidal stability provided by the PEI polymer, that keeps the aggregation limited and also provides a thick surface layer that minimizes the inter-particle dipolar interactions within each cluster. In this way, Néel relaxation of the magnetic moments are not affected by agglomeration, keeping the SPA values constant in time. These characteristics make these MNPs appropriate for in vitro and in vivo applications of magnetic hyperthermia, were reproducibility is a key factor.

## Experimental Section

All samples studied in this work consisted on PEI-coated MNPs in water, synthetized according the modified oxidative hydrolysis method described in detail elsewhere.[7, 49] In short, iron (II) sulphate heptahydrate (FeSO$_4$·7 H$_2$O), sodium hydroxide (NaOH), potassium nitrate (KNO$_3$), sulphuric acid (H$_2$SO$_4$) and polyethylenimine (PEI, M$_W$ = 25 kDa) were mixed in a three-necked flask bubbled with N$_2$. Firstly we prepared a solution with NAOH and KNO$_3$ in the three-necked flask, and at the same time a solution containing H$_2$SO$_4$, FeSO$_4$ ·7H$_2$O and PEI, which is added dropwise under constant stirring. The synthesis process was held at 90ºC for 24 hours under N$_2$, when the time is over, PEI-MNPs was separated by magnetic decantation and washed several time with deionized water to get physiological pH.[7]

The physicochemical properties of the resulting MNPs-PEI size and their morphology were analysed by transmission electron microscopy (TEM) using a FEI Tecnai T20 microscope and operating at 200 keV; and high resolution transmission electron microscopic (HR-TEM) images were obtained by using a FEI Tecnai F30 microscope operated at an acceleration voltage of 300 KV. Sample was prepared laying a drop of a dilute suspension of magnetite nanoparticles in water on a carbon-coated copper grid and allowing the solvent to evaporate at room temperature. Colloidal properties of the samples were measured in a 90 Plus Particle Size Analyser (Brookhaven Instrument Corp., USA), the hydrodynamic properties of the nanoparticles in dilute suspensions using a solution 0.01 M KCl at physiological pH as dilution carrier. The Fe concentration was determined by VIS-UV transmission spectrophotometry (Shimadzu UV-160) using thiocyanate complexation through the following reaction [50]

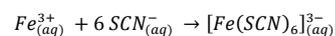

$$Fe^{3+}_{(aq)} + 6\, SCN^-_{(aq)} \rightarrow [Fe(SCN)_6]^{3-}_{(aq)}$$

Before complexation, the MNPs were dissolved in HCl 6 M-HNO$_3$ (65%) at 50-60 ºC during 2 h. Potassium thiocyanate was then added to the Fe$^{3+}$ solution to form the iron-thiocyanate complex, which has strong absorbance at wavelength $\lambda$=478 nm. The iron concentration was determined by comparing the sample absorbance to a calibration curve.

The study of specific absorption rate was performed in a commercial magnetic field applicator (DM1 applicator, nB Nanoscale Biomagnetics, Spain) using magnetic fields from 3,98 to 23,87 kA/m at several frequencies from 260 to 832 kHz. Calorimetric experiments were conducted exposing 1 ml of the magnetic colloid in a vacuum-insulated Dewar connected to a vacuum pump ($10^{-7}$ mbar). A fiber-optic measuring probe was placed at the centre of the colloid to sense the temperature of the sample. A 'dead time' of 2-5 minutes was allowed before each experiment in order to reach thermal equilibrium of the samples.

## Acknowledgements

This work was supported by the Spanish Ministerio de Economia y Competitividad (MINECO, projects MAT2010-19326, PRI-PIBAR-2011-1384 and MAT2013-42551). Technical support from LMA-INA and SAI-UZ is also acknowledged.

**Keywords:** Magnetic nanoparticles • hyperthermia • colloidal stability • surface modification • Box-Lucas fit

**Entry for the Table of Contents** (Please choose one layout)

Layout 1:

# FULL PAPER

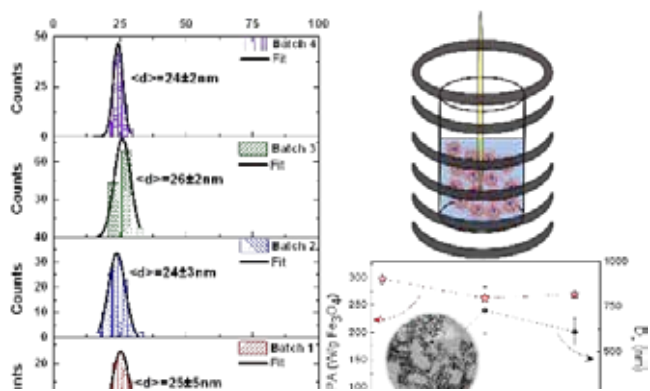

**Key Topic: Magnetic Hyperthermia**

*B. Sanz\*, M.P. Calatayud, N. Cassinelli, M.R. Ibarra and G.F. Goya\**

*E-mail: beasanz@unizar.es, goya@unizar.es.*

*Page No. – Page No.*

**Title**

Layout 2:

# FULL PAPER

((Insert TOC Graphic here; max. width: 11.5 cm; max. height: 2.5 cm; NOTE: the final letter height should not be less than 2 mm.))

**Key Topic: Magnetic Hyperthermia**

*B. Sanz\*, M.P. Calatayud, N. Cassinelli, M.R. Ibarra and G.F. Goya\**

*E-mail: beasanz@unizar.es, goya@unizar.es.*

*Page No. – Page No.*

**Title**

Text for Table of Contents

\*one or two words that highlight the emphasis of the paper or the field of the study